\documentclass[aps,prl,twocolumn,showpacs,superscriptaddress]{revtex4-1}

\usepackage{graphicx}
\usepackage{epsfig}
\usepackage{epstopdf}
\usepackage{bm}
\usepackage{color}

\begin{document}

\def\sr327{Sr$_3$Ru$_2$O$_7$}
\def\327{Sr$_3$Ru$_2$O$_7$}
\title{Study of the electronic nematic phase of \sr327 with precise control of the applied magnetic field vector}

\author{J. A. N. Bruin}

\affiliation{SUPA, School of Physics and Astronomy, University of St Andrews, St Andrews KY16\ 9SS, United Kingdom}

\author{R. A. Borzi}

\affiliation{Instituto de Investigaciones Fisicoqu\'\i{}micas
  Te\'oricas y Aplicadas, UNLP-CONICET, 1900 La Plata, Argentina}

\author{S. A. Grigera}
\affiliation{SUPA, School of Physics and Astronomy, University of St Andrews, St Andrews KY16\ 9SS, United Kingdom}
\affiliation{Instituto de F\'{\i}sica de L\'{\i}quidos y Sistemas Biol\'ogicos, UNLP-CONICET, 1900 La Plata, Argentina}

\author{A. W. Rost}

\affiliation{SUPA, School of Physics and Astronomy, University of St Andrews, St Andrews KY16\ 9SS, United Kingdom}

\author{R. S. Perry}

\affiliation{SUPA, School of Physics, University of Edinburgh, Edinburgh EH9 3JZ, United Kingdom}

\author{A. P. Mackenzie}

\affiliation{SUPA, School of Physics and Astronomy, University of St Andrews, St Andrews KY16\ 9SS, United Kingdom}
\affiliation{Max-Planck Institute for Chemical Physics of Solids, D-01187 Dresden, Germany}

\date{\today}

\begin{abstract}
We report a study of the magnetoresistivity of high purity \sr327, in the vicinity of its electronic nematic phase. By employing a triple-axis (9/1/1T) vector magnet, we were able to precisely tune both the magnitude and direction of the in-plane component of the magnetic field ($H_\parallel$). We report the dependence of the resistively determined anisotropy on $H_\parallel$ in the phase, as well as across the wider temperature-field region. Our measurements reveal a high-temperature anisotropy which mimics the behaviour of fluctuations from the underlying quantum critical point, and suggest the existence of a more complicated phase diagram than previously reported.
\end{abstract}

\maketitle

Since the pioneering work of Pomeranchuk \cite{pomeranchuk}, it has been known that a fluid of interacting electrons might, in principle, develop a spontaneous macroscopic anisotropy. Through the 1990s, solutions of idealised models of electrons on lattices showed that a variety of spin and charge orders can occur that break the symmetries of the underlying lattice and are purely driven by electron-electron interactions. The most celebrated of these, stripes, have been widely discussed in relation to the cuprate high temperature superconductors \cite{kivelson2003}. When the stripes are static there is both a macroscopic anisotropy and an observable translational symmetry breaking, but states in which translational symmetry is preserved may also exist. The analogy between this behaviour and that seen in liquid crystals has led to the adoption of liquid crystal terminology in correlated electron systems: a static stripe phase is equivalent to an electronic smectic, while a metal showing rotational but not translational symmetry breaking is often referred to as an electronic nematic \cite{kivelson1998}.

In recent years, rapid advances have been made. Electronic nematicity has been reported in high purity two dimensional electron gases (2DEGs) \cite{lilly1999,pan1999}, \sr327 \cite{borzi2007}, cuprates \cite{ando2002,hinkov2008,daou2010,lawler2010}, pnictides \cite{chuang2010,chu2010,kasahara2012} and URu$_2$Si$_2$ \cite{okazaki2011}, stimulating a considerable body of theoretical work \cite{fradkin2010}. In some cases, however, interpretation of the observations is complicated by the coexistence of nematicity with other ordering phenomena. One of the main requirements for further progress is identifying systems in which the onset of anisotropy can be unambiguously determined, and their physical properties established in depth. The material that is the subject of this paper, \sr327, is a particularly attractive host for such experiments. It is a layered perovskite metal in which conduction occurs in stacks of Ru-O bilayers. The two Ru-O planes in each bilayer couple relatively strongly, but bilayer-bilayer coupling is weak, so the electronic structure is nearly two-dimensional \cite{ikeda2000,singh2001,tamai2008,mercure2010}. In high magnetic fields of approximately 8 tesla applied normal to the Ru-O planes, a novel phase forms in the purest crystals \cite{grigera2004,rost2009}. Tilting the field to a small angle, $\theta$, produces an in-plane field component which uncovers a large two-fold in-plane transport anisotropy whose easy and hard directions are perpendicular and parallel, respectively, to the direction of the in-plane field \cite{borzi2007}. 

While neutron studies were unable to resolve any accompanying anisotropic response of the lattice to the onset of the electronic anisotropy \cite{borzi2007}, it was recently shown by dilatometry measurements to be less than one part in 10$^{6}$ \cite{stingl2011}. This suggests that the coupling of the electronic distortion to the crystal lattice is extremely weak, and means that \sr327 offers, in principle, an ideal opportunity to study the way in which electronic nematics form and behave. In this paper, we report an experimental study of the transport-derived nematic susceptibility which reveals the full richness of the \sr327 phase diagram.

In order to investigate the effect of an in-plane field with high experimental precision, it is necessary to tune its magnitude and orientation with respect to the sample. This cannot be done directly using a standard single-axis rotator. Although interesting experiments can be attempted using octagonal samples with multiple contact configurations \cite{borzi2011}, changing contact configuration during an experiment introduces geometric correction factors that can be difficult to determine with accuracy. A precise investigation of transitions from two- to four-fold transport requires two axis rotation by some means. The standard approach would be to construct a two-axis mechanical rotator, but performing the rotation inevitably introduces heat which disrupts measurements at dilution refrigerator temperatures. The experiments described here result from a different experimental approach, using a bespoke triple-axis vector magnet capable of producing 9/1/1 tesla along $z$, $x$ and $y$. We studied carefully characterised single crystals cut either into ‘needle’ shapes (2 x 0.3 x 0.2 mm$^3$ typical dimensions) or octagonal plates (1.8 x 1.8 x 0.1 mm$^3$ typical dimensions). Transport measurements were performed in a dilution refrigerator using four-probe a.c. techniques employing room temperature passive amplification to achieve voltage noise levels of 100pV/$\sqrt{\textrm{Hz}}$. Careful centring using the vector field capability allowed $\theta$ to be determined to an accuracy of 0.1$^{\circ}$ relative to the crystallographic $c$ axis \cite{janthesis}.

\begin{figure}
\centerline{\includegraphics[width=0.85\columnwidth]{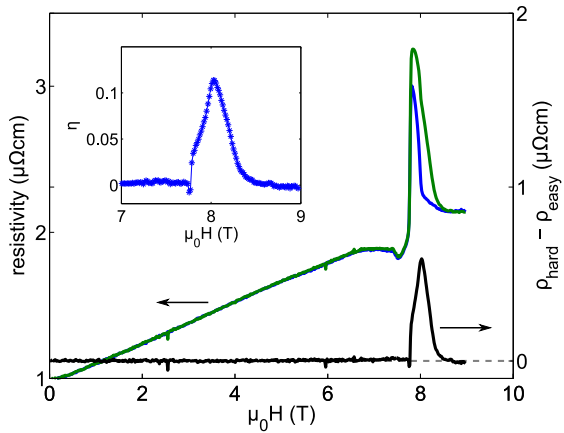}}
\caption{Magnetoresistance of \sr327 versus an applied field tilted at $\theta$ = 6.4$^{\circ}$ parallel (green trace) and perpendicular (blue trace) to a current applied along the tetragonal $a$ axis, as well as the difference between these (black trace). The inset shows the dimensionless anisotropy $\eta$ (see: main text). Up to 7.78 tesla it is zero within our experimental resolution but then jumps sharply into a region of strong transport anisotropy which weakens gradually before returning to zero above approximately 8.5 tesla.}
\label{fig1}
\end{figure}

In Fig. \ref{fig1} we show the magnetoresistivity of a single crystal of \sr327 as a function of the applied field modulus at $\theta$ = 6.4$^{\circ}$ and $T$ = 100 mK. At this tilt angle the in-plane field $H_{\parallel}$ reaches 1 tesla when the $c$ axis field is 9 tesla. In both measurements the current was passed along the tetragonal $a$ axis \footnote{In the unfolded tetragonal zone the $a$ axis is along the Ru-Ru bond direction.}, but for the blue curve the in-plane field component is perpendicular to the current (the ‘easy’ direction), while for the green curve it is parallel to it (the ‘hard’ direction). Both the current and voltage contacts to the crystal remain identical, so this is a precise measurement of whether the resistivity changes for the two perpendicular in-plane field directions or not. Within our resistivity noise level of 2 n$\Omega$cm/$\sqrt{\textrm{Hz}}$, the resistivity is identical until the sharp jump at 7.78 tesla, as demonstrated by the black curve ($\rho_{hard}-\rho_{easy}$) in Fig. \ref{fig1}. At higher fields there is a clear resistive anisotropy, which then falls to zero within experimental resolution near the experimentally limited maximum field. A convenient dimensionless measure of the degree of anisotropy is the ratio $\eta = (\rho_{hard} - \rho_{easy}) /(\rho_{hard} + \rho_{easy})$. We illustrate the onset and subsequent disappearance of anisotropy by plotting $\eta$ as a function of field in the inset of Fig. \ref{fig1}.

The data shown in Fig. \ref{fig1} emphasise the first key result of this paper: the uniqueness of \sr327 as a model material for the investigation of electronic nematicity. In all the systems in which large anisotropies have been observed, a symmetry-breaking field has been required to reveal them. In the 2DEGs, sample strain has been suggested as the source of this symmetry-breaking field \cite{fradkin2010}, while in the cuprates and pnictides it arises from a pre-existing lattice anisotropy. In the latter case, deconvolving the nematicity due to electronic interactions from coupling to the anisotropic lattice is challenging, though ways of doing so have been proposed \cite{ando2002,chuang2010,chu2012}. In \sr327, two issues could in principle complicate the interpretation of transport data. Firstly, there is one report of a small lattice anisotropy of approximately 5 parts in 10$^4$ even at zero applied magnetic field \cite{kiyanagi2004}. Although small, and not reported in other crystallographic studies \cite{shaked2000jssc,hu2010}, it could be a source of two-fold transport anisotropy. Secondly, an in-plane field component could lead to transport anisotropy because of the difference between transverse and longitudinal magneto-resistance \footnote{If a lattice anisotropy of 5 parts in 10$^4$ creates no observable transport anisotropy, a change over two orders of magnitude less is not expected to do so either, while any difference between transverse and longitudinal magnetoresistance is a strong inverse function of the scattering rate, which is higher in the anisotropic region than outside it.}. The fact that we resolve two-fold transport anisotropy only in a restricted range of fields shows that both effects are negligible in \sr327. Combined with the dilatometry data showing that the lattice anisotropy introduced by entering the region of two-fold transport is less than 1 part in 10$^6$, this provides strong evidence that the two-fold transport in \sr327 is induced by electron interactions. Further, the fact that the magnetoresistance makes such well-defined changes from four-fold ($\eta = 0$ to within high experimental precision) to two-fold ($\eta > 0$) strongly suggests that it is associated with symmetry breaking of the interacting electron system. 

\begin{figure}
\centerline{\includegraphics[width=0.7\columnwidth]{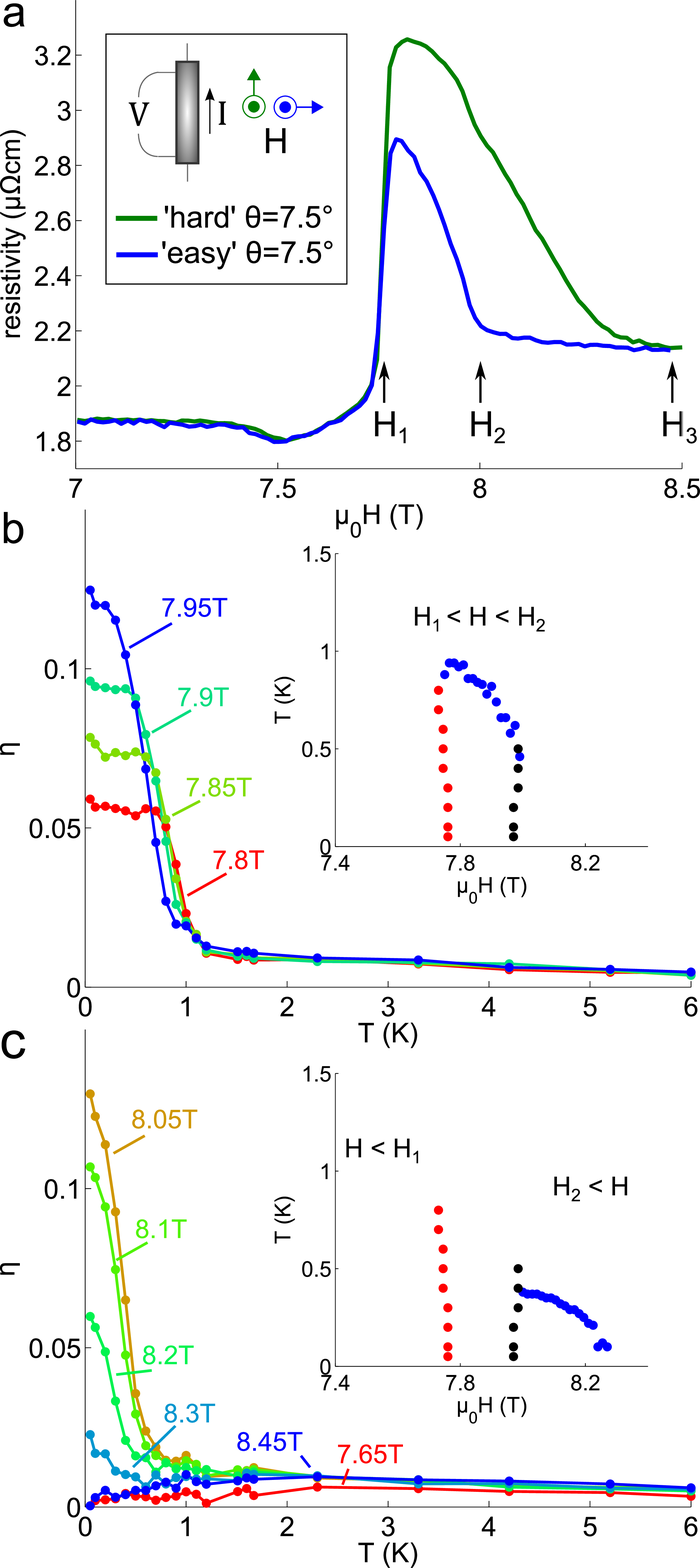}}
\caption{(a) data for magnetoresistance along the hard (green) and easy (blue) directions as a function of field tilted to $\theta$ = 7.5$^{\circ}$, $T$ = 100 mK. The labels $H_1$ and $H_2$ mark the fields at which peaks in the magnetic susceptibility have been identified (see main text), taking into account the weak $\theta$ dependence of these features \cite{raghu2009}. (b,c) Dimensionless anisotropy $\eta$ (see: main text) as a function of temperature for $H_1 < H < H_2$ and $H_2 < H < H_3$, respectively. The data (solid markers) are interpolated from field sweeps at fixed temperature. The insets present points of greatest slope in $\eta(T)$ (blue markers), as well as the loci of $H_1$ (red markers) and $H_2$ (black markers), extracted as the points of greatest slope in $\rho_{hard}(H)$.}
\label{fig2}
\end{figure}

In Fig. \ref{fig2}a we show data in the close vicinity of the region of resistive anisotropy, for $\theta$ = 7.5$^{\circ}$ and $T$ = 100 mK. The field labels facilitate comparison with previous thermodynamic and magnetic data, which have established the existence of first order phase transitions at $H_1$ and $H_2$ on the basis of dissipative peaks in the imaginary part of low-frequency a.c. susceptibility \cite{perry2004} and features in the magnetocaloric effect \cite{rost2009}. The data show that, as the field increases, the initial sharp rise in the anisotropy occurs at $H_1$. Although there is a feature in $\rho_{hard}$ at $H_2$, anisotropy does not disappear there, but persists to the higher field $H_3$. 

The existence of resistive anisotropy between $H_2$ and $H_3$, beyond the boundaries of the previously-identified phase, is the second key result reported in this paper. Electrical transport in \sr327 is based mainly on bands with Ru 4$d_{xy}$, 4$d_{xz}$ and 4$d_{yz}$ character. A body of theoretical work has shown that electronic anisotropy can arise from instabilities of either the quasi-1D 4$d_{xz,yz}$-based bands \cite{raghu2009,lee2009} or from a (bilayer-split) quasi-2D model \cite{puetter2007,yamase2009}. The real Fermi surface is quite complicated, however, and the portions which are most strongly renormalised have hybrid 4$d_{xy}$ and 4$d_{xz,yz}$ character \cite{tamai2008}. It seems plausible, therefore, that multiple instabilities may exist, as addressed recently in a model predicting a multi-component phase diagram \cite{puetter2011}, and evidence for lattice constant anisotropy above $H_2$ has also been reported \cite{stingl2012}. 

The observed transport anisotropy between $H_2$ and $H_3$ shows that the phase diagram of \sr327 is more complex than previously thought, motivating a thorough investigation of the field, temperature and $\theta$ dependence of the anisotropy. We have performed such a study, measuring $\eta$ in the vector magnet for 7T $< B <$ 8.5T, 0.1 K $< T <$ 6K and $\theta$ = 1$^{\circ}$, 2$^{\circ}$, 4$^{\circ}$, 6.4$^{\circ}$ and 7.5$^{\circ}$. In Figs. \ref{fig2}b and \ref{fig2}c we show samples from this data set, namely the temperature dependence of $\eta$ at $\theta$ = 7.5$^{\circ}$, for temperatures between 6 K and 100 mK, for fields between the $H_1$ and $H_2$, and fields above $H_2$ and below $H_1$ respectively.

The data in Fig. \ref{fig2}b are relevant to an important issue in the study of electronic nematicity: why anisotropy is observed only in the presence of a symmetry-breaking field. The most likely reason is the inherent degeneracy between the order developing along the $a$ or $b$ axes of a square-planar system. If a phase transition to a nematic state occurs, it would be expected to be accompanied by the formation of domains which would mask the microscopic anisotropy from observation in macroscopic measurements \cite{fradkin2010}. In layered compounds like \sr327 it is also possible that even in mono-domain samples the preferred direction might rotate from layer to layer, again preventing the observation of anisotropy \cite{puetter2007,yamase2009}. If a microscopic order parameter exists, the field that is applied to reveal the anisotropy must couple to it, so what is actually measured is the response of the system to the application of that field, which in the case of \sr327 is $H_{\parallel}$. The first feature of Fig. 2b that is consistent with this basic picture is the pronounced high temperature `tail' seen in $\eta$. Small but resolvable anisotropy exists to 6 K, the maximum temperature shown. Below approximately 1 K, $\eta$ rises rapidly from its `background' value. As shown in the inset (blue markers), the temperatures at which this happens agree closely with those of second order phase transitions previously identified for $H_{\parallel}$ = 0 \cite{rost2009}. The finite high temperature anisotropy along with a pronounced increase at 1K are characteristic of the behaviour expected for an ordered system in the presence of an external field to which it couples, and are reminiscent of previous measurements on the anisotropic state that exists at high field in high purity 2DEGs \cite{lilly1999}.

Data for $H_2 < H < H_3$ (Fig. \ref{fig2}c) share some but not all of the features of the data shown in Fig. \ref{fig2}b. The `tail' extending to 6 K is still evident, as is the rise in $\eta$ at low temperatures. The temperatures at which this rise occurs are lower and there is no evidence for the low-temperature saturation of $\eta$ seen at most fields between $H_1$ and $H_2$, but the data close to $H_2$ show little qualitative difference either side of it. This suggests that there might be a second phase between $H_2$ and $H_3$, raising the question of why it has not so far been observed in thermodynamic measurements. The issue may be disorder. The previously identified phase between $H_1$ and $H_2$ is strongly dependent on sample purity, and only becomes well-defined in samples with mean free paths of several thousand angstroms. If there were a phase between $H_2$ and $H_3$ with a lower transition temperature, it might be expected to suffer significant disorder broadening even at this high level of sample purity, making a thermodynamic signature difficult to observe. Our transport observations motivate further thermodynamic experiments to clarify this issue. 

The third key finding from our study of the field and temperature dependence of $\eta$ is also contained in Fig. \ref{fig2}c. The 7.65T and 8.45T curves (red and blue markers) are representative of the temperature dependence of $\eta$ at fields away from, but in close proximity to, the main regions of anisotropy, both for $H < H_1$ and $H > H_3$. At high temperatures $\eta$ is similar to the values measured for $H_1 < H < H_3$. Instead of rising at low temperatures, however, it drops towards zero. This is characteristic of a quantum critical system tuned away from its critical field. At high temperatures, critical fluctuations affect a wide range of field, but this range is cut off when the temperature is too low to populate them. There is extensive thermodynamic and transport evidence that, away from its critical field, \sr327 behaves like a quantum critical system (with the actual approach to the quantum critical point cut off by the phase formation), but until this experiment there was no evidence that the fluctuations had any nematic character. The nematic response that we report here is small, and would have been difficult to determine with certainty without the use of a vector magnetic field. It is therefore not clear whether the primary fluctuations of the system are nematic or whether the nematicity is a by-product of more dominant fluctuations of some other kind, but it is intriguing to see a nematic signal with such direct characteristics of quantum criticality. It will be interesting to see if it can be enhanced by the application of in-plane stress, to allow the phase diagram to be traced out in detail. 

\begin{figure}
\centerline{\includegraphics[width=0.95\columnwidth]{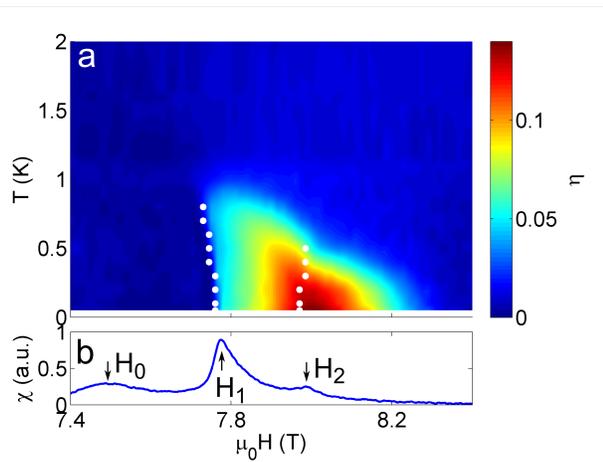}}
\caption{(a) the temperature and field dependence of the dimensionless anisotropy ratio $\eta$ measured, as described in the main text, in a field tilted at $\theta$ = 7.5$^{\circ}$. The white dots show the loci of $H_1$ and $H_2$, as extracted from points of greatest slope of $\rho(H)$. The endpoints of the first order lines were determined by comparison to a.c. magnetic susceptibility. (b) the field dependence of the a.c. magnetic susceptibility at 100mK, $\theta \approx 7^{\circ}$, over the same range of magnetic fields (background subtracted).}
\label{fig3}
\end{figure}

In Fig. \ref{fig3}a we summarise all the data in the vicinity of the regions of strong anisotropy in a colour scale contour plot of $\eta$ for $\theta$ = 7.5$^{\circ}$, based on interpolating data taken at 17 closely spaced temperatures. The data shown in the insets of Fig. \ref{fig2}b and \ref{fig2}c for $H_1$ and $H_2$ are superposed here as white dots. The contouring of $\eta$ in the higher field region is suggestive of that region being associated with a second phase. 

Comparison with magnetic susceptibility data for similar $\theta$ (Fig. \ref{fig3}b) reveals a final noteworthy feature of the phase diagram. In addition to the sharp peaks at $H_1$ and $H_2$, which are known to signal first order phase transitions \cite{grigera2003,grigera2004,rost2009}, there is a broader peak at the lower field $H_0$. Although it is associated with a larger moment change than that at $H_2$, there is no evidence from either the imaginary part of the magnetic susceptibility or the magnetocaloric effect that it signals a first order magnetic transition. It might in principle be a crossover or a broadened second order transition (either a critical end-point or a symmetry-breaking transition to finite $q$ order of some kind). Whether it is a crossover or a phase transition, it is clearly not associated with the onset of low temperature transport anisotropy. The resistivity falls slightly, but retains four-fold symmetry within our experimental resolution. In contrast, the disappearance of anisotropy at $H_3$ does not coincide with any large feature in the magnetic susceptibility. The overall picture that emerges from these measurements is that the phase diagram of \sr327 may be richer than previously thought. High precision thermodynamic measurements extending to temperatures below 100 mK are highly desirable. 

In conclusion, we have used a vector magnetic field to perform a series of high-precision measurements of anisotropic transport in \sr327. Within our high experimental resolution, we observe well-defined transitions from four-fold to two-fold to four-fold symmetry of in-plane resistivity as a function of magnetic field. Transport anisotropy exists in a region of field higher than that bounding a previously identified anisotropic phase. Even when the resistivity is four-fold symmetric at low temperatures, some anisotropy can be induced at higher temperatures, consistent with nematic fluctuations being an integral feature of the broader quantum critical phase diagram that is known to exist for \sr327.

We thank S. A. Kivelson and S. Raghu for stimulating discussions. This work was supported by the EPSRC (UK), RAB and SAG were partially supported by the Royal Society (UK), CONICET and ANPCyT (Argentina), and APM holds a Royal Society - Wolfson Research Merit Award.

\end{document}